\documentclass[aps,prl,epsfig,twocolumn,showpacs,preprintnumbers,amsmath,amssymb,superscriptaddress,floatfix]{revtex4}
\usepackage{epsfig}
\usepackage{dcolumn}
\usepackage{bm}
\usepackage{braket}
\usepackage[latin1]{inputenc}                                 
\usepackage{color}                                            
\usepackage{calc}                                             
\usepackage{hhline}                                           

\begin{document}

\title{Valence bond glass on an fcc lattice in the
  double perovskite {Ba$_2$YMoO$_6$}.}
\author{M.~A.~de~Vries} \email{m.a.devries@physics.org}
\affiliation{School of Physics and Astronomy, E. C. Stoner Laboratory,
  University of Leeds, Leeds. LS2 9JT, UK}
\affiliation{School of Physics \& Astronomy,
  University of St-Andrews, the North Haugh, KY16 9SS, UK}
\author{A.~C.~Mclaughlin}
\affiliation{Department of Chemistry, University of Aberdeen, Meston
  Walk, Aberdeen AB24 3UE, UK}
\author{J.-W.~G. Bos}
\affiliation{School of
  Chemistry, University of Edinburgh, King's Buildings, Mayfield
  Road, Edinburgh EH9 3JZ, UK} 
\affiliation{Department of Chemistry - EPS, Heriot-Watt
University, Edinburgh, EH14 4AS, UK}

\begin{abstract}
We report on the unconventional magnetism in the cubic B-site
ordered double perovskite Ba$_2$YMoO$_6$, using AC and DC magnetic
susceptibility, heat capacity and $\mu$SR. No magnetic order is
observed down to 2~K while the Weiss temperature is $\sim
-160$~K. This is ascribed to the geometric frustration in the
lattice of edge-sharing tetrahedra with orbitally degenerate Mo$^{5+}$
$s=1/2$ spins. Our experimental results point to a gradual freezing of
the spins into a disordered pattern of spin-singlets, quenching the
orbital degeneracy while leaving the global cubic symmetry
unaffected, and providing a rare example of a valence bond glass. 
\end{abstract}

\date{\today}

\maketitle{}  

Magnetic insulators with lattices in which antiferromagnetic (AF)
bonds are geometrically frustrated have been studied widely in the
pursuit of exotic quantum ground states such as spin liquid~\cite{
  Greedan:GFM, Lhuillier:04}. Such non-classical ground states have
mainly been sought in low dimensional structures such as the triangular
lattice system
$\kappa$-(BEDT-TTF)$_2$Cu$_2$(CN)$_3$~\cite{Shimizu:03} and the kagome
antiferromagnet herbertsmithite~\cite{Shores:05}. Materials
with a geometrically frustrated face
centred cubic (fcc) lattice have in this respect received much less
attention. The 12 near-neighbour magnetic bonds
$J_1$ between the [000] and [$\frac{1}{2}\frac{1}{2}0$] spins on the
fcc lattice form a network of edge-sharing tetrahedra (Fig.  1). When
these bonds are AF ($J_1>0$) the magnetism is geometrically
frustrated, giving rise to a large (but not macroscopically
large~\cite{Lines:65} as for the kagome lattice) ground-state
manifold of spin configurations unrelated by symmetry. Further
neighbour interactions ($J_2$) along the 6 [100] vectors lift this
degeneracy only partially; $J_2 < 0$ (along the 6 [100] vectors) leads to
type I order, weak AF exchange ($0<J_2<2J_1$) to type III order and
stronger AF exchange $J_2>2J_1$ to type II order. Thermal/quantum
fluctuations and quenched disorder have been shown to result in a bias for
respectively collinear and anti collinear states within
these degenerate ground state manifolds~\cite{Yamamoto:72, Henley:87,
  Yildirim:98}, an entropic selection effect termed ``order from
disorder''~\cite{Villain:80}. This is in agreement with experiments on
well-known compounds of rock-salt structure such as
MnO~\cite{Shull:51,Goodwin:04},
Cd$_{1-x}$Mn$_x$Te~\cite{Galazka:80}, NiO,
MnSe~\cite{Shull:51}. Classical type I, II or III order has also been
confirmed for $s=1/2$~\cite{Oguchi:85, Lefmann:01,
  Zhang:02} although less is known about the physics at the boundaries
between the classical phases. In
this Letter we describe the unconventional magnetism in the
compound Ba$_2$YMoO$_6$, providing experimental evidence that
an exotic valence bond glass (VBG)~\cite{Chamon:05,Tarzia:08} state
can stabilize at the boundary between the known classical phases on
the fcc lattice. Such a disordered state has been predicted to be possible
even in the absence of structural disorder, as an example of a
non-equilibrium quantum ground state~\cite{Chamon:05}. 

\begin{figure}[ht]
\epsfig{file=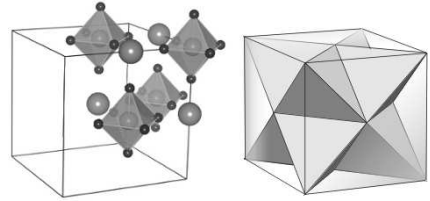, width = 2.2in}
\caption
{Four MoO$_6$ octahedra (shaded grey) and Y ions (large spheres) in
  the cubic unit cell of Ba$_2$YMoO$_6$ (left). The Mo$^{5+}$ ions form a
  lattice of edge-sharing tetrahedra (right). The cubic lattice
  constant is 8.389~\AA.}
\label{figure:bymoview}
\end{figure}

The B-site ordered double perovskites are of general stoichiometry
$A_2BB'$O$_6$ where the $A$ site typically hosts alkaline-earths and
lanthanides and the $B$ sites can host $3d$, $4d$ and $5d$ transition
metal (TM) ions. Depending on the combination of $B$ and $B'$ site
ions, electronic phases from strongly correlated metals via
half-metals~\cite{Kobayashi:98} and
semiconductors~\cite{bos_structural_2004, bos_crystal_2005} to Mott
insulating can be realized. Mott insulating $4d$ and $5d$ TM
compounds are rare. The occurrence of this insulating phase in the double
perovskites is due to the large distance between the TM ions, of the
order of 5 to 6~\AA. Examples of Mott insulators are
Ba$_2$LaRuO$_6$ and Ca$_2$LaRuO$_6$~\cite{Battle:83}, respectively
type III and type I antiferromagnets. Sr$_2$CaReO$_6$~\cite{Wiebe:02}
and Sr$_2$MgReO~\cite{Wiebe:03} (the Re$^{6+}$ has $s=1/2$) have
spin-glass ground states, consistent with a negligible $J_2$ along the
pathway Re-O-$B'$-O-Re. There is a large group of Mo$^{5+}$ compounds
Ba$_2Ln$MoO$_6$ with $Ln = $~Nd, Sm, Eu, Gd, Dy, Er, Yb and
Y~\cite{Cussen:06}. The Mo$^{5+}$ has a singly occupied $4d$ $t_{2g}$
level with $s=1/2$. Due to the strong spin-orbit coupling in
$4d$ TM ions in a cubic crystal field this is expected to lead to
a $j=3/2$ triplet~\cite{Goodenough:68,
  Erickson:07}. Only the larger
lanthanide compounds ($Ln$ = Nd, Sm and Eu) have N\'eel order (Type I,
implying $J_2 <0$) coinciding with a weak Jahn-Teller
distortion~\cite{Cussen:06, ACM:06, ACM:08}, while the exchange
interaction is of the order of 100~K~\cite{Cussen:06}.  The other
compounds were found to be paramagnetic and with cubic
symmetry at all temperatures. Ba$_2$YMoO$_6$
is the simplest of these compounds because the Y$^{3+}$ ion does
not carry a magnetic moment. The magnetic exchange is mainly
via the 90$^{\circ}$ B'-O-O-B' $\pi - \pi$ bonds~\cite{Battle:83,Cussen:06},
giving rise to 12 near-neighbour AF $J_1$ bonds for each spin, across
the edges of the tetrahedra (Fig. 1).

Polycrystalline Ba$_2$YMoO$_6$ was prepared by the solid state
reaction of stoichiometric oxides of Y$_2$O$_3$, MoO$_3$ and
BaCO$_3$ powders of at least 99.99\% purity. These were ground,
die-pressed into a pellet and heated under flowing 5\%
H$_2$/N$_2$. The final synthesis temperature was
1200-1250$^{\circ}$C with three intermediate regrinding steps to
ensure phase homogeneity. It was found that a first heating step
of $\sim 2$~hr at 900$^{\circ}$C in air and thorough
homogenization helps to prevent the formation of BaMoO$_4$ and
Y$_2$O$_3$ impurities. Phase purity was confirmed by laboratory
X-ray powder diffraction.
In a related paper~\cite{ACM:09} neutron powder diffraction
results are discussed, which show that the Y/Mo site disorder is
less than 1\%. The diamagnetic analog, Ba$_2$YNbO$_6$, was
prepared at 1200$^{\circ}$C in air from YNbO$_4$ and BaCO$_3$.
The sample magnetisation was measured on a Quantum Design
magnetic property measurement system (MPMS) in fields up to 5~T. The heat capacity
was measured on a Quantum Design physical
property measurement system (PPMS), using $7.0$~mg of a
sintered pellet. The $\mu$SR experiment was carried out at MUSR at
ISIS, UK.        

\begin{figure}[htbp]
\epsfig{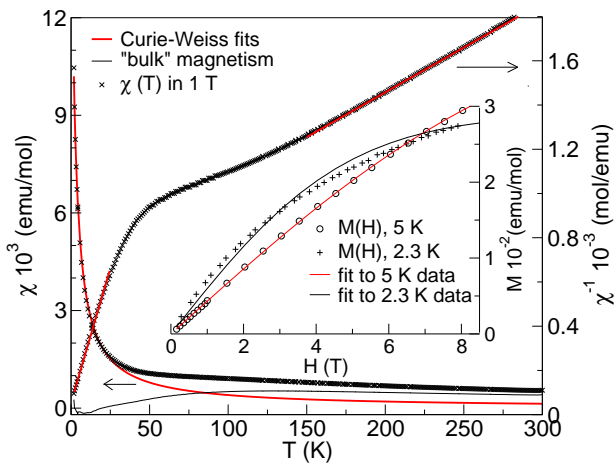}
\caption
{(color online) The DC magnetic susceptibility $\chi$ (\sf{x}, left axis)
  and $\chi^{-1}$ (\sf{x}, right axis) of Ba$_2$YMoO$_6$ measured in
  1~T. Curie-Weiss fits in the two linear regimes in
  $\chi^{-1}$ are indicated in grey (red) and the black line gives the
  difference between the total susceptibility and the low-temperature
  Curie term. The inset shows $M(H)$ at 2.3 (+) and 5 K ($\circ$) and fits to
  the data with Brillouin functions, accounting for 7\% of the Mo $s=1/2$
  spins at 5K but only for 2\% at 2.3 K.}
\label{figure:dcsusc}
\end{figure}

The DC magnetic susceptibility measured in a 1 T field is shown in
Fig.~\ref{figure:dcsusc}. A
Curie-Weiss fit to the high temperature susceptibility yields a
Weiss temperature of -160~K and a Curie constant of 0.25 emu
mol$^{-1}$ K$^{-1}$, small compared to the 0.38 emu mol$^{-1}$ K$^{-1}$
expected for $s=1/2$. This difference is attributed to strong quantum
fluctuations common in low-spin antiferromagnets and previously observed in
double perovskites~\cite{Erickson:07}. Below 25~K a second linear
regime is observed in $\chi^{-1}$, corresponding (for a 1~T field) to
a $\sim 10$\% fraction of all the $s=1/2$ moments (or $\sim 5$\% if they
have the full $j=3/2$ where $g_J=4/3$) and a Weiss temperature of $-2.3$~K indicating
weak AF exchange. This fraction is too large to be ascribed
directly to either structural disorder or an impurity phase in the
sample. Furthermore, fits to $M(H)$ measured at 2.3 and 5~K (inset of
Fig.~\ref{figure:dcsusc}) with Brillouin functions lead to estimates
of respectively 2 and 7\% of all spins, compared to 10\% for fits to
the $M(T)$ curve. This suggests that the apparently quasi-free spins are
an emergent property of the (disorder free) system. 

\begin{figure}[ht]
\epsfig{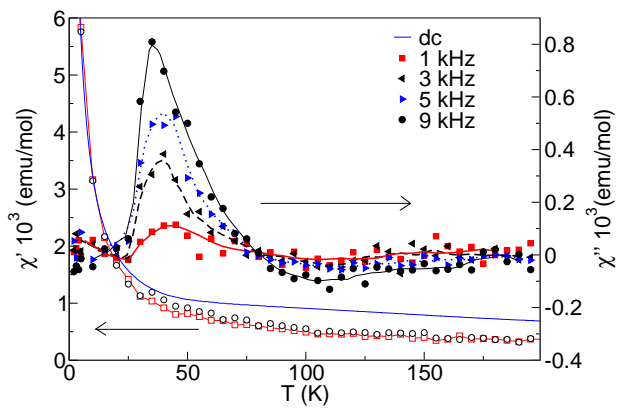}
\caption
{(Color online) The temperature dependence of the dispersive (left axis, open
  symbols) and dissipative (right axis, filled symbols) components of
  the AC susceptibility. The solid lines are guides to the eye.}
\label{figure:acsusc}
\end{figure}

The AC susceptibility measured with a field amplitude of 5 Oe and zero
DC offset field is shown in 
Fig.~\ref{figure:acsusc}. The dispersive part of the AC
susceptibility ($\chi '$) is almost frequency independent and is
comparable to the diverging low temperature DC susceptibility.  The
dissipative part ($\chi ''$) shows a frequency dependent maximum
between 26 and 70 K. Remarkably, the maximum gradually gets sharper as
the frequency increases, instead of weaker as expected for a
spin-glass transition. The agreement between the DC susceptibility and
$\chi '$ below 20~K is a strong indication that the Curie-term can be
ascribed to the weakly coupled spins.

\begin{figure}[ht]
\epsfig{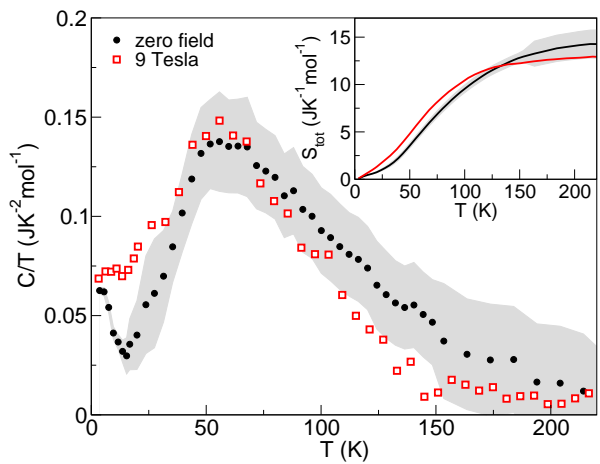}
\caption
{(Color online) The magnetic heat capacity obtained by
  subtracting the heat capacity of the diamagnetic analogue
  Ba$_2$YNbO$_2$ in zero field (black dots) and in 9 Tesla (open
  squares). The experimental error for the 9~T data is comparable to
  that indicated for the zero field data (grey area). The inset shows
  the total entropy release as a function of temperature in zero field
  (black line) and in 9~T (red line).}
\label{figure:heatcap}
\end{figure}

The heat capacity associated with the single Mo$^{5+}$ $4d$ electron
in Ba$_2$YMoO$_6$ (as shown in Fig.~\ref{figure:heatcap}) was obtained
from comparison with the heat capacity of the diamagnetic analogue
Ba$_2$YNbO$_6$. The heat capacity from phonons of Ba$_2$YMoO$_6$ is
expected to be lower than for Ba$_2$YNbO$_6$ by a factor 0.991 due to
the mass difference between the Mo and Nb nuclei. However this is
small compared to the experimental error in
the sample mass which is known with 10\%
accuracy. For this reason the heat capacity was measured well into the
paramagnetic regime and matched to the heat capacity of Ba$_2$YNbO$_6$
above 200~K (150~K) for the zero-field (9~T) measurements~\cite{footnoteHc}. 
The magnetic entropy is gradually released over a wide
range of temperatures with a broad maximum around 50~K. No anomalies
corresponding to phase transitions are observed, only a gradual
freezing, quenching all degrees of freedom associated with the
orbitally degenerate $t_{2g}$ $s=1/2$ $4d$ electrons.  As shown in the
inset of Fig.~\ref{figure:heatcap}, the total entropy recovered 
$S_{\text{tot}}=12\pm 2$ JK$^{-1}$mol$^{-1}$, close to the $R\ln 4 = 11.5$ expected
for a $j = 3/2$ quadruplet (the $j=l-s=1/2$ doublet lies at much
higher energies~\cite{Goodenough:68,Erickson:07}). Below 25~K only
$\sim 5$\% of the entropy is released, in agreement with the Curie fit
to the low temperature susceptibility which was found to correspond to
$\sim 5$\% of the Mo$^{5+}$ if these remaining spins have $j=3/2$. In a
9~T magnetic field most of the magnetic entropy shifts to lower temperatures.

\begin{figure}[ht]
\epsfig{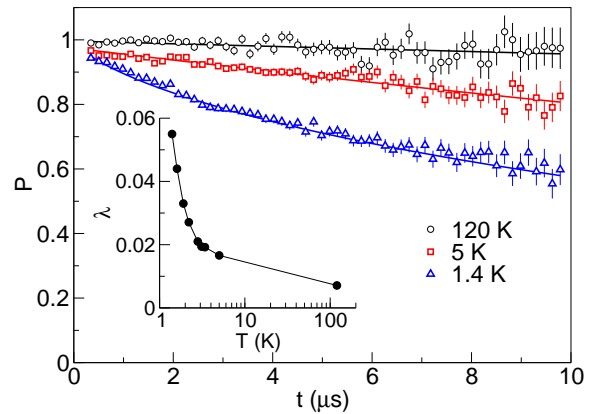}
\caption
{(Color online) The muon spin relaxation at 120, 5 and 1.4
  K. The relaxation follows an exponential decay
  ($P(t)=\exp[-t/\lambda]^{\beta}$, solid lines) with $\beta=1$ for all
  but the 1.4 K data, where $\beta = 0.7$. The
  temperature dependence of the relaxation rate $\lambda$
  is shown in the inset.}
\label{figure:musr}
\end{figure}

To gain a better understanding of the gradual freezing and the
appearance of apparently weakly-coupled spins a $\mu$SR experiment was
carried out. The zero-field muon spin relaxation spectra at 120 K, 5 K
and 1.4 K are shown in Fig.~\ref{figure:musr}. There is no evidence of
muon relaxation due to nuclear spins which confirms that the main
muon stopping site is near the O$^{2-}$ ions. At 120~K there is no
muon relaxation, as expected for a paramagnetic state. Remarkably, at
5~K a muon relaxation is still only just detectable. If the maximum in
the AC susceptibility is due to a conventional spin-glass transition a
Lorentzian Kubo-Toyabe muon relaxation is expected below the spin
glass transition, as observed in the related system
Sr$_2$MgReO$_6$~\cite{Wiebe:03}. The very slow muon relaxation
observed at 5~K in Ba$_2$YMoO$_6$ indicates there are no static
moments. At the same time the heat capacity data shows that at 5~K
most of the magnetic entropy associated with $j=3/2$ is quenched,
implying static order. The majority of spins must therefore have bound
into (non-magnetic) static spin-singlet ``valence bonds'' in which
also the orbital degrees of freedom are quenched. The
moderate increase in the muon relaxation rate below 5~K is then due to
slowing-down of a small fraction of the spins which are left isolated
as domain walls or defects in a (disordered) valence bond crystal
(VBC). The best characterization of this state is probably a valence
bond glass (VBG) as described in Ref.~\onlinecite{Tarzia:08}. 

The magnetic properties of Ba$_2$YMoO$_6$ are very different to
those of the related compound Sr$_2$MgReO$_6$~\cite{Wiebe:03}, where a
first order transition to a conventional spin glass state is
observed. That the crossover in Ba$_2$YMoO$_6$ is not a conventional
spin-glass transition is also clear from the unusual frequency
dependence of the AC susceptibility. The gradual freezing and
cross-over region around 50~K are consistent with a pseudogap
predicted for the VBG~\cite{Tarzia:08}. This gap, which corresponds to
a spin-singlet dimerization energy scale, is filled by levels
corresponding to emergent weakly-coupled spins which give rise to a
diverging susceptibility as the temperature is decreased. In close
agreement with Ref.~\onlinecite{Tarzia:08} the observed low
temperature susceptibility follows a power law $\chi \propto
(T-T_s)^{-\gamma}$ with $\gamma \approx 1$. As noted earlier, this
contribution from effectively weakly-coupled spins can not be related
one-to-one to any structural disorder but arises as a cooperative
effect, due to the amorphous arrangement of spin-singlets. The heat
capacity does not become zero at the lowest temperature measured which
is consistent with a small residual entropy and an ungapped spectrum
as expected for the VBG. 

The classical ground-state energy is highest when $J_2 \approx 2J_1$,
at the cross-over between type III ($J_2<2J_1$) and type II magnetism
(the energy per spin is $-J_1/2+J_2/4$ for $J_2\leq 2J_1$). One
possibility is that around this cross-over
a spin-singlet state is energetically favored. A complete
explanation of why spin-singlets stabilize will in the present case
also involve the orbital degrees of freedom; It is the valence-bond
formation rather than a Jahn-Teller effect that fixes the orbital
orientation. This could be accompanied by local structural distortions
which do not lead to experimentally
observable~\cite{Cussen:06,ACM:09} structural changes because the
valence bonds do not form a regular pattern within the crystal. As
first observed by Goodenough and Battle~\cite{Battle:83}, the band
width is an important factor in understanding the magnetism in double
perovskites with $4d$ and $5d$ transition metal ions. Because
Ba$_2$YMoO$_6$ is a relatively wide-band insulator it could also be that
the stability of a spin-singet state can be understood in terms of the $t-J$
model~\cite{Dagotto:91}. 

In conclusion, the B-site ordered double perovskite Ba$_2$YMoO$_6$ has
Mo$^{5+}$ ions with a singly-occupied degenerate $t_{2g}$ orbital in a
cubic crystal field and with $s=1/2$. These moments are located on an
fcc lattice and coupled antiferromagnetically, with a Weiss
temperature of $-160$~K. At  high temperature the single-ionic $j=3/2$
moments are strongly reduced by quantum fluctuations, consistent with
the formation of stable spin-singlet dimers at low
temperatures. Remarkably, the dimerization pattern appears to be
disordered, giving rise to emergent effectively unpaired spins with a
diverging susceptibility as the temperature is reduced, in agreement
with theoretical predictions of a VBG~\cite{Tarzia:08}.  It is
proposed that the stabilisation of spin-singlets is a result of the
fine-tuning of the ratio $J_2/J_1$ to the boundary between the
classical type II and  type III phases. Clearly the present results
provide a strong motivation for further theoretical explorations of
the phase diagram of the fcc antiferromagnet and of the interplay
between spin and orbital degrees of freedom in $4d$ and $5d$
transition metal compounds. 
This is the first observation of a VBG in a quantum magnet. 
Further experimental studies are needed
to understand the relationship between the structural and magnetic
disorder in this class of materials. 

The Royal Society of Edinburgh (JWGB) and Leverhulme Trust (ACM)
are acknowledged for financial support. C.~L. Henley and
H.~M. R\o{}nnow are gratefully
acknowledged for fruitful discussions. S.~J. Ray, P. King and P.
Baker are gratefully acknowledged for assistance with the $\mu$SR experiments.


\begin{thebibliography}{10}
\bibitem{footnoteHc}
The data thus obtained gives a lower limit to the heat capacity
associated with the singly occupied $4d$ $t_{2g}$ level, which
is sufficient for the present purpose.
\end{thebibliography}

\end{document}